\documentclass[10pt, onecolumn, conference]{IEEEtran}

\usepackage[T1]{fontenc}
\usepackage[utf8]{inputenc}
\usepackage[english]{babel}
\usepackage{array}
\usepackage{graphicx}
\usepackage{booktabs}
\usepackage{amsmath}
\usepackage{amssymb}
\usepackage{multirow}
\usepackage{makecell}
\usepackage{pifont}
\usepackage{caption}
\usepackage{xcolor}
\usepackage{url}
\usepackage{float}  % para [H]
\usepackage{breakcites}
\usepackage[hidelinks]{hyperref}
\usepackage{orcidlink}

\providecommand{\doi}[1]{\url{https://doi.org/#1}}

\graphicspath{{figures/}}

\makeatletter\g@addto@macro\UrlBreaks{\do\-\do\/\do\.\do\_}\makeatother
\newcommand{\cmark}{\textcolor{green!60!black}{\ding{51}}}
\newcommand{\xmark}{\textcolor{red!70!black}{\ding{55}}}

\newcommand{\sysname}{ChimangoScan}

\begin{document}

\title{Vulnerabilities, Secrets and Misconfiguration in the\\
Highest-Exposure Docker Hub Images}

\author{
\IEEEauthorblockN{Cristhian Kapelinski\orcidlink{0009-0005-5750-022X}\IEEEauthorrefmark{1},
Beatriz Machado\orcidlink{0009-0002-2750-0323}\IEEEauthorrefmark{1},
Diego Kreutz\orcidlink{0000-0003-0830-0238}\IEEEauthorrefmark{1}}
\IEEEauthorblockA{\IEEEauthorrefmark{1}AI Horizon Labs, Federal University of Pampa (UNIPAMPA), Alegrete, Brazil}
\IEEEauthorblockA{\{cristhianavila, beatrizmachado\}.aluno@unipampa.edu.br, diegokreutz@unipampa.edu.br}
}

\maketitle

\begin{abstract}
Docker Hub underpins containerized software, yet large-scale security measurements of its images are infrequent and rely on a single scanner. We present \sysname{}, a pipeline that crawls Docker Hub and ranks images by a layer-graph \emph{exposure} score: its pull count plus those of every image inheriting its layers. From a \textbf{12.7~million}-repository crawl we scanned the \textbf{52{,}895} highest-exposure images, covering \textbf{84.7\%} of all 663.8~billion recorded pulls, with \emph{six} open-source scanners (\textbf{170.4~million} findings). Known vulnerabilities are near-universal. \textbf{96.3\%} of images carry one, \textbf{93.4\%} carry a critical one, and \textbf{98.0\%} carry a misconfiguration. Yet the reported posture depends heavily on the scanner. Of the distinct vulnerabilities the three vulnerability scanners find, only \textbf{2.7\%} are reported by all three and 66.8\% by one. Secrets show the opposite pattern. TruffleHog flags 76.9\% of images, yet \textbf{99.7\%} of its hits are non-credentials by hand-labeling. A lone scanner's count is thus largely an artifact of that scanner.
\end{abstract}


\section{Introduction}
\label{sec:intro}

Software supply chain attacks are projected to cost USD~60~billion in 2025, up from USD~46~billion in 2023; Gartner estimated 45\% of organizations would face one by 2025, and over 700{,}000 malicious open-source packages have now been cataloged.\footnote{\href{https://cybersecurityventures.com/software-supply-chain-attacks-to-cost-the-world-60-billion-by-2025/}{Cybersecurity Ventures, 2023}, \href{https://www.sonatype.com/state-of-the-software-supply-chain/2024/scale}{Sonatype, 2024}} Containers sit on that supply chain, and Docker Hub is the registry underneath: 92\% of IT professionals report using containers, and over half of organizations run most or all of their applications in containers.\footnote{\href{https://www.docker.com/blog/2025-docker-state-of-app-dev/}{Docker, 2025 survey}, \href{https://www.cncf.io/wp-content/uploads/2025/04/cncf_annual_survey24_031225a.pdf}{CNCF, 2024}} Docker Hub hosts over 14~million images and advertises ``11B+ monthly image downloads'';\footnote{\href{https://www.docker.com/products/docker-hub/}{Docker Hub}} cumulatively, our crawl sums \textbf{663{,}779{,}362{,}551} historical pulls (about 663.8~billion) across \textbf{12{,}716{,}568} public repositories (Section~\ref{sec:stage1}). When a developer writes \texttt{FROM}~\texttt{python:3.12}, what runs in production is pulled unmodified from an image someone else published, and a flaw in that base is inherited by every image built on it. The reach can be enormous; \texttt{alpine} draws \textbf{11.8~billion} pulls of its own, but images built on its layers add roughly \textbf{71~billion} more, so one critical vulnerability in \texttt{alpine} reaches about \textbf{83~billion} inherited pulls.

This parent-to-child propagation, first documented by Shu et al.~\cite{shu2017dockerhub}, makes the registry a software-supply-chain attack surface~\cite{ladisa2023sok} whose reach varies widely. A critical flaw in \texttt{alpine:latest} reaches the whole ecosystem, while the same flaw in an unused leaf image reaches almost nothing. \sysname{} captures this with an \emph{exposure} score (Section~\ref{sec:exposure}). Since images also leak credentials and ship insecure defaults, a faithful measurement must look beyond CVE (Common Vulnerabilities and Exposures) counts. Yet prior ecosystem-scale measurements each rely on a \emph{single} detector: Shu on Clair~\cite{shu2017dockerhub}, Liu on one bespoke tool~\cite{liu2020understanding}, Dahlmanns on one secret-detection pipeline~\cite{dahlmanns2023secrets}, and Dr.\ Docker on one SCA (Software Composition Analysis) engine over 3{,}000 of its 33{,}952 images~\cite{shi2025drdocker}. Each count thus carries unquantified tool-dependence. The studies that \emph{do} compare scanners use small samples (Kaur et al., a pairwise Jaccard overlap of 0.59 to 0.80, on a 0-to-1 scale where 1 is identical, over 44 images~\cite{kaur2021scientific}; Churakova et al., best pair 0.69 over 48 images~\cite{churakova2025vex}; Javed and Toor, 59 images~\cite{javed2021quality}; Mills et al., 380 images~\cite{mills2023longitudinal}). No prior work crosses the two axes at ecosystem scale.

This paper addresses that gap with \textbf{\sysname{}}, a pipeline that crawls the Docker Hub namespace, reconstructs the image layer graph, ranks images by downstream exposure, and scans the highest-exposure ones with six independent scanners. Its contributions are \emph{(i)}~the \emph{exposure} score $E(I)$ (Section~\ref{sec:exposure}), the first to fold an image's own pull count and those of its entire downstream subtree into one scalar; \emph{(ii)}~a six-scanner measurement of all \textbf{52{,}895} top-ranked repositories (170.4\,M findings), two orders of magnitude more images than prior multi-scanner studies (Section~\ref{sec:res-prevalence}); \emph{(iii)}~an ecosystem-scale measurement of inter-scanner disagreement (Section~\ref{sec:res-divergence}), showing that the posture a single tool reports is substantially an artifact of that tool (only \textbf{2.7\%} of distinct vulnerabilities are flagged by all three vulnerability scanners, and \textbf{99.7\%} of TruffleHog's hits, found in 76.9\% of images, are non-credentials by hand-labeling); and \emph{(iv)}~a public dataset (Section~\ref{sec:dataset}) released with the pipeline (metadata for 12{,}716{,}568 repositories, the layer graph, and the multi-scanner reports for the 52{,}895 scanned images).

\section{Related Work}
\label{sec:related}

Table~\ref{tab:related} positions \sysname{} against three lines of prior work: large-scale security measurements of Docker Hub, studies of inter-scanner agreement, and studies of secrets and misconfiguration in container images.

\begin{table}[!htpb]
\centering
\caption{Positioning against prior work.}
\label{tab:related}
\renewcommand{\arraystretch}{0.98}
\footnotesize
\setlength{\tabcolsep}{2.5pt}
\begin{tabular*}{\textwidth}{@{\extracolsep{\fill}}l c r c c c c c}
\toprule
\textbf{Work} & \textbf{Year} & \textbf{Images} & \textbf{Scanners} & \textbf{Vuln.} & \textbf{Misconfig.} & \textbf{Secrets} & \makecell{\textbf{Cross-}\\\textbf{scanner}} \\
\midrule
Shu et al.~\cite{shu2017dockerhub}          & 2017 & 356{,}218    & 1 & \cmark & \xmark & \xmark & \xmark \\
Zerouali et al.~\cite{zerouali2019outdated} & 2019 & 7{,}380      & 1 & \cmark & \xmark & \xmark & \xmark \\
Liu et al.~\cite{liu2020understanding}      & 2020 & 2{,}227{,}244 & 1 & \cmark & \xmark & \xmark & \xmark \\
Wist et al.~\cite{wist2021vulnerability}    & 2021 & 2{,}500      & 1 & \cmark & \xmark & \xmark & \xmark \\
Dahlmanns et al.~\cite{dahlmanns2023secrets} & 2023 & 337{,}171   & 1 & \xmark & \xmark & \cmark & \xmark \\
Dr.\ Docker~\cite{shi2025drdocker} & 2025 & 33{,}952 & 1 & \cmark & \cmark & \cmark & \xmark \\
\midrule
Kaur et al.~\cite{kaur2021scientific}       & 2021 & 44           & 4 & \cmark & \xmark & \xmark & \cmark \\
Mills et al.~\cite{mills2023longitudinal}   & 2023 & 380          & 6 & \cmark & \xmark & \xmark & \cmark \\
\midrule
\textbf{\sysname{}} (this work)           & \textbf{2026} & \textbf{52{,}895} & \textbf{6} & \cmark & \cmark & \cmark & \cmark \\
\bottomrule
\end{tabular*}
\end{table}

\noindent\textbf{Large-scale Docker Hub measurements.} Docker Hub has been measured periodically since Shu et al.~\cite{shu2017dockerhub}, whose DIVA tool scanned 356{,}218 images with Clair, found $>$180 vulnerabilities/image, and traced parent-to-child propagation. Liu et al.~\cite{liu2020understanding} scaled to 2.2~million images and found 42 malicious ones, and Zerouali et al.~\cite{zerouali2019outdated} measured technical lag (how far a package trails its latest fixed version) in 7{,}380 Debian-based images. All three used a \emph{single} detector. A complementary line reconstructs ecosystem \emph{structure}: Opdebeeck et al.~\cite{opdebeeck2023inheritance} build an inheritance network from 636{,}625 images, the closest prior work to our layer graph but at a fraction of the scale and without multi-scanner measurement. \textbf{The closest work.} Dr.\ Docker~\cite{shi2025drdocker} crawls \textbf{12{,}079{,}309} repositories, \textbf{637{,}259} fewer than our \textbf{12{,}716{,}568}, builds an ancestry-hashed layer graph, selects \emph{influential} images, and scans 33{,}952 images for five threat classes. \sysname{} adopts two of its \emph{ideas}, the cap-refining keyword crawl and the ancestry-hashed layer identifier, and forks its open Go codebase (Section~\ref{sec:pipeline}): we implement Stage~I, an unimplemented stub upstream, and re-engineer Stage~II for distributed, resumable operation. It departs in two respects: \emph{(i)~scale and tooling}: Dr.\ Docker scans only 3{,}000 sampled images for vulnerabilities with one SCA engine, whereas \sysname{} runs \emph{six} scanners on all scanned repositories; \emph{(ii)~reproducibility}: its malware verdicts use a corporate antivirus, whereas \sysname{} uses only open-source scanners. \textbf{Further related lines.} Scanner agreement has been measured only on small samples (four scanners, 44 images, Jaccard 0.59 to 0.80~\cite{kaur2021scientific}; six scanners, 380 images~\cite{mills2023longitudinal}). We revisit this question at the scale of our corpus. For secrets and misconfiguration, Dahlmanns et al.~\cite{dahlmanns2023secrets} found secrets in 8.5\% of images and over 52{,}000 private keys, building on repository secret-leakage work~\cite{meli2019gitleaks}. Rosa et al.~\cite{rosa2022fixing} and Cito et al.~\cite{cito2017empirical} study Dockerfile smells, and Ahamed et al.~\cite{ahamed2021audit} derive audit use cases that align image checks of this kind (content trust, \texttt{HEALTHCHECK}, non-root user) with the NIST SP~800-190 and OWASP container-security guidelines. \sysname{} covers secrets (TruffleHog) and misconfiguration (Dockle) in the same pipeline.

\section{The \sysname{} Measurement Pipeline}
\label{sec:pipeline}

\begin{figure}[!htpb]
    \centering
    \includegraphics[width=\linewidth]{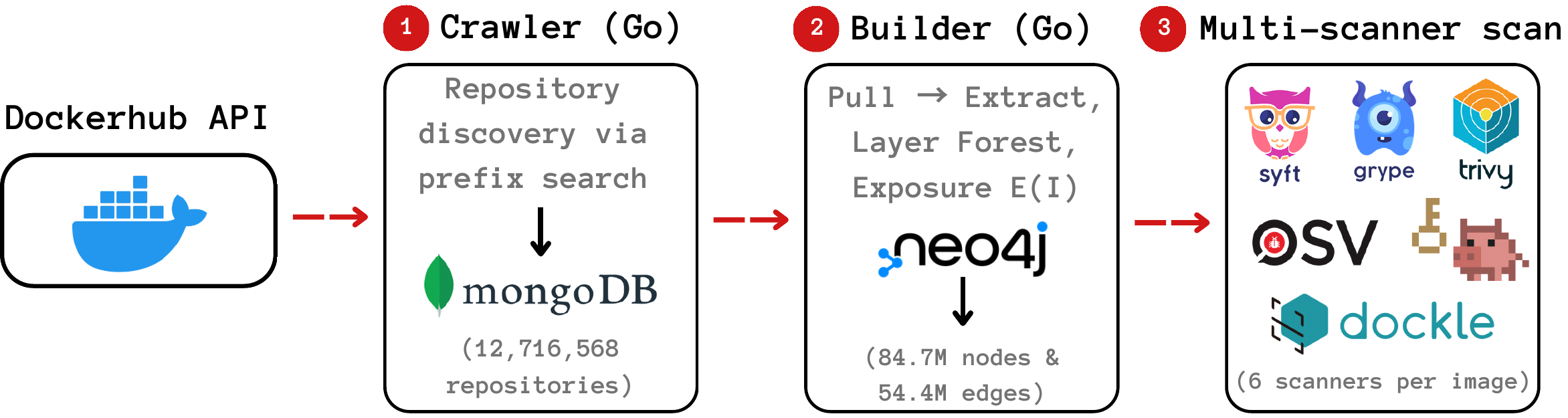}
    \caption{The three-stage pipeline.}
    \label{fig:pipeline}
\end{figure}

\sysname{} is a three-stage pipeline (Figure~\ref{fig:pipeline}). Stage~I discovers Docker Hub repositories; Stage~II resolves each into concrete images and reconstructs the image \emph{layer graph}; Stage~III scans images, prioritized by the layer graph, with six independent scanners. Stages~I and~II are written in Go, Stage~III in Python. Two methodological ideas are adopted from Dr.\ Docker~\cite{shi2025drdocker} and credited where used: the keyword crawl that recursively refines any query saturating the API's result cap (Stage~I) and the ancestry-hashed layer-node identifier (Stage~II). We fork Dr.\ Docker's open Go codebase, implementing the crawl (a stub upstream) ourselves and re-engineering the layer-node builder for distributed operation (Sections~\ref{sec:stage1} and~\ref{sec:stage2}). The new pipeline elements are the \emph{exposure-based prioritization} of the scan queue (Section~\ref{sec:exposure}) and the \emph{six-scanner consolidation pipeline} of Stage~III (Section~\ref{sec:stage3}).

\paragraph{Terminology.} We use four distinct nouns throughout. A \emph{repository} is a namespaced bucket on Docker Hub (e.g.\ \texttt{library/alpine}); the Hub reports a single pull count per repository. A \emph{tag} (e.g.\ \texttt{latest}, \texttt{3.18}) is a mutable label inside a repository pointing to a specific image. An \emph{image} is identified by a content-addressed \texttt{sha256} digest; two repositories or tags may share a digest when byte-identical, and every aggregate statistic in this paper deduplicates by image digest. A \emph{layer} is one filesystem slice inside an image, identified by its own digest and shared across images built from the same base; the \texttt{IS\_BASE\_OF} layer graph of Section~\ref{sec:stage2} captures that sharing. Stage~I enumerated \textbf{12{,}716{,}568} repositories; Stage~II resolved the most-pulled repositories (the \emph{popularity head}) in decreasing pull-count order, down to roughly \textbf{72} pulls (\textbf{44.05\%} of the crawl, Section~\ref{sec:stage2}), so the layer graph and the downstream-propagation counts built on it (Section~\ref{sec:res-reach}) are \emph{lower bounds} over the full namespace.

\subsection{Stage~I: Repository Discovery}
\label{sec:stage1}

Stage~I enumerates the Docker Hub namespace through the public Search API. Because the API caps any query at 10{,}000 results, a crawler must refine any keyword whose query saturates the cap (the cap-refining idea of Dr.\ Docker~\cite{shi2025drdocker}); we realize this as a \emph{prefix-trie traversal} of repository names, seeding 38 single-character prefixes (\texttt{a} to \texttt{z}, \texttt{0} to \texttt{9}, \texttt{-}, \texttt{\_}) and expanding any prefix that returns the 10{,}000-result ceiling into its 38 children (\texttt{py}~$\rightarrow$~\texttt{pya}, \texttt{pyb}, \dots), shortest-first. The crawler is a distributed, resumable Go system and ran over \emph{several months} in 2026. One caveat bounds exhaustiveness: the API's ElasticSearch backend treats the hyphen as a token separator, so repository names with a hyphen in the middle are explored only best-effort; we are not aware of repositories missed this way but cannot rule it out.

By scan time the crawler had indexed \textbf{12{,}716{,}568} repositories from \textbf{2{,}051{,}801} prefix queries (all in \textbf{2026}), recording each one's pull count, stars, and last-updated timestamp. Docker advertises ``14M+ container images''; our 12.7~million \emph{repositories} (each holding many tags) cover very nearly the entire public namespace, since every cap-saturated prefix was expanded. The \textbf{663.8~billion} pulls of Table~\ref{tab:pulldist} are a \emph{cumulative historical} total, not Docker's ``11B+ monthly''.

The pull-count distribution (reproducing Dr.\ Docker~\cite{shi2025drdocker}) is extremely heavy-tailed: a median of \textbf{62} pulls but a 99th percentile of 20{,}781 and a maximum of $2.4\times10^{10}$, with 113 repositories drawing \textbf{60.6\%} of all pulls and 95\% pulled fewer than a thousand times (Table~\ref{tab:pulldist}\footnote{The distribution covers the \textbf{12{,}680{,}591} repositories with a recorded pull count; \textbf{35{,}977} of the 12{,}716{,}568 crawled lack one.}). The layer graph shows the same concentration structurally: only a small minority are \emph{base images}, with heavy-tailed downstream counts. A uniform sample would thus spend almost all its effort on images the ecosystem barely uses; the exposure score (Section~\ref{sec:exposure}) exploits exactly this imbalance.

\begin{table}[!htpb]
\centering
\caption{Repository pull-count distribution over the crawl.}
\label{tab:pulldist}
\renewcommand{\arraystretch}{0.95}
\small
% single 6-column table: two halves as columns of one table; 7 buckets plus the
% total split 4/4, so both sides have the same number of rows (no empty cell).
\begin{tabular}{@{}lrr@{\hspace{2.6em}}lrr@{}}
\toprule
\textbf{Pull count} & \textbf{Repos} & \textbf{\% pulls} & \textbf{Pull count} & \textbf{Repos} & \textbf{\% pulls} \\
\midrule
$\geq$1\,B    & 113      & 60.6\% & 100\,k--1\,M   & 42{,}145                & 2.1\%    \\
100\,M--1\,B  & 461      & 20.1\% & 1\,k--100\,k   & 572{,}662               & 0.8\%    \\
10\,M--100\,M & 2{,}595  & 11.1\% & $<$1\,k        & 12{,}050{,}759          & $<$0.1\% \\
1\,M--10\,M   & 11{,}856 & 5.3\%  & \textbf{Total} & \textbf{12{,}680{,}591} & \textbf{100\%} \\
\bottomrule
\end{tabular}
\end{table}

\subsection{Stage~II: Image Resolution and the Layer Graph}
\label{sec:stage2}

A repository is not an image: it is a set of tags, each resolving to one or more architecture-specific image digests. The Stage~II builder claims each repository exclusively from the crawl pool (so none is processed twice), queries the Docker Hub API for each one's tags and manifests, then pulls each image and extracts the ordered list of filesystem \emph{layers} that compose it. The builder consumes repositories in \emph{decreasing pull-count order}, so the popularity head is processed first; at the dataset freeze used in this paper Stage~II has resolved \textbf{5{,}601{,}045} of the 12.7\,M crawled repositories (44.05\%) into \textbf{6{,}399{,}608 tags} and \textbf{7{,}416{,}671 image digests}, with the remaining tail of rarely pulled repositories still pending.

Layers are the unit of reuse: an image built \texttt{FROM}~\texttt{python:3.12} physically shares that base's layers, and \sysname{} reconstructs this sharing as a directed graph. For the layer-node identifier we adopt Dr.\ Docker's ancestry-hashed \emph{scheme}~\cite{shi2025drdocker}, instantiated with SHA-256: $\mathrm{id}=\mathrm{sha256}(\mathrm{parent\_id}\,\|\,\mathrm{sha256}(\mathrm{layer\_digest}))$, so the identifier encodes the layer's full ancestry and identical-content layers with different histories stay distinct. Our builder is a distributed Go service: each worker atomically claims the next unbuilt repository from a shared MongoDB pool (resumable after crashes, stale claims released), pulls its recent tags and manifests, extracts the ordered layer list, and inserts \texttt{IS\_BASE\_OF} edges from each layer to the one above it. Repositories deleted between crawl and build are retired rather than stalling the queue. The deterministic identifier gives each layer at most one parent, so the graph is a \emph{forest of out-trees} (disjoint trees with all edges pointing away from the root), stored in Neo4j with \textbf{54{,}382{,}383} \texttt{IS\_BASE\_OF} edges over 84.7\,M nodes. A flaw in a reused base layer is therefore inherited by its whole subtree. The exposure prioritization on this forest (Section~\ref{sec:exposure}) is specific to this work: Dr.\ Docker ranks by dependency weight, its count of downstream images, whereas \sysname{} folds an image's own pull count and its downstream reach into the single score $E(I)$.

\subsection{Exposure-Based Prioritization}
\label{sec:exposure}

Scanning all 6.7\,M images is infeasible, and they are not equally worth scanning: a flaw in a base image thousands inherit risks far more than one in a rarely pulled leaf. \sysname{} ranks the queue by an \emph{exposure} score that rewards downstream reuse. Prior work ranks bases by descendant \emph{count}~\cite{shu2017dockerhub} or by dependency-graph weight~\cite{shi2025drdocker}; exposure instead folds an image's own popularity and that of its entire downstream subtree into a single scalar (the container-layer analog of dependency-graph impact scoring~\cite{ruan2025vpss}), linearly ordering the queue by total ecosystem reach. For an image $I$, write $R_I$ for the repository publishing $I$, $p(R)$ for the pull count of a repository $R$, $L_I$ for $I$'s top layer, and $\mathcal{D}(L_I)$ for the set of images whose top layer is a \emph{strict} descendant of $L_I$ in the forest; a descendant image $D \in \mathcal{D}(L_I)$ is published by repository $R_D$. Because bases nest, we attribute each image to a single \emph{owner}, the image of greatest own pull count among it and its ancestors, so no downstream pull is double-counted. Exposure is then $I$'s own pull count plus those of every image it owns:

\begin{equation}
E(I) \;=\; p(R_I) \;+\!\!\! \sum_{\substack{D \,\in\, \mathcal{D}(L_I)\\[1pt] \mathrm{owner}(D)\,=\,I}} p(R_D). \label{eq:exposure}
\end{equation}

The queue is sorted by $E(I)$, with ties broken by the downstream image count $|\mathcal{D}(L_I)|$ (Dr.\ Docker's dependency weight~\cite{shi2025drdocker}). Single-owner attribution credits each downstream pull to the most-pulled base in its lineage, and a daemon recomputes the scores in a near-linear pass as the crawl grows, placing \texttt{alpine:latest} first ($E\approx8.3\times10^{10}$). Two rules guard against misplaced credit. A byte-identical republication (a bare \texttt{FROM}) is credited to the \emph{canonical} most-pulled reference, and a generic metadata-only layer (\texttt{LABEL}, \texttt{CMD}) recurring under many parents is barred from owning downstream images; both rules only move credit toward the more-pulled base. Because Stage~II resolves only recent tags, a subtree's owner is occasionally a less-pulled repository pinning the same base; this occurs only among the least-pulled repositories Stage~II resolved. The ranking head has two kinds of image. General-purpose bases (\texttt{alpine}, \texttt{ubuntu}, \texttt{debian}) draw 72\% to 86\% of their exposure from downstream images. Heavily pulled application images (\texttt{nginx}, \texttt{memcached}, \texttt{busybox}, Istio) are seldom used as a base, so their exposure is almost all their own pulls.

\subsection{Stage~III: The Multi-Scanner Scan Pipeline}
\label{sec:stage3}

Stage~III serves the exposure-ordered queue to workers that pull each image \emph{pinned by digest} and run \emph{six} static scanners covering four dimensions: software inventory (\textbf{Syft}); package vulnerabilities from three databases (\textbf{Trivy}, \textbf{Grype}, \textbf{OSV-Scanner}, over the National Vulnerability Database (NVD)/OSS-Index, the Anchore VulnDB, and Google's Open Source Vulnerabilities (OSV) database respectively, so each detects CVEs the others miss); misconfiguration against the CIS (Center for Internet Security) Docker Benchmark (\textbf{Dockle}); and hardcoded secrets across 700+ detectors (\textbf{TruffleHog}). The queue is keyed by each repository's \texttt{latest} tag, so the reported posture is one of \texttt{latest} tags, not all published tags. A per-scanner adapter normalizes each output to a common six-level severity scale; TruffleHog secrets are marked critical only when verified live, otherwise \emph{medium}. Findings are then \emph{consolidated} per image but \emph{not} deduplicated across scanners, so a CVE found by three scanners counts three times. That non-deduplication is what makes inter-scanner agreement measurable (Section~\ref{sec:res-divergence}).

\subsection{Dataset}
\label{sec:dataset}

The scan ran on 13 commodity x86-64 workers (Debian~12 or Ubuntu~24.04, 8 to 32 threads, 32 to 128\,GB of RAM, no GPU), each scanning \texttt{linux/amd64} images. The unit is the \emph{repository reference}, one \texttt{latest} image per repository; the campaign scanned the \textbf{52{,}895} highest-exposure ones (\textbf{84.7\%} of all \textbf{663.8~billion} recorded pulls), and deduplicating by content digest to \textbf{51{,}751} distinct images leaves the results essentially unchanged. The reports total \textbf{170{,}373{,}044} findings (141.7\,M package vulnerabilities, 23.2\,M inventory components, 4.96\,M secrets, 0.58\,M misconfigurations; Trivy's built-in secret and misconfiguration checks feed the latter two categories alongside TruffleHog and Dockle); we release the full dataset with the pipeline (Table~\ref{tab:dataset}).

\begin{table}[!htpb]
\centering
\caption{Dataset summary.}
\label{tab:dataset}
\renewcommand{\arraystretch}{0.95}
\setlength{\tabcolsep}{4pt}
\footnotesize
% the two halves sit side by side in a tabular sized to their content, and the
% merged-findings block is centered on its own below, so neither stretches the
% other across the full onecolumn text width.
\begin{tabular}{@{}c@{\hspace{0.06\textwidth}}c@{}}
\begin{tabular}[t]{@{}lr@{}}
\multicolumn{2}{@{}l}{\textbf{Crawl and layer graph}}\\
\midrule
Repositories crawled              & 12{,}716{,}568 \\
Repositories Stage~II-resolved    & 5{,}601{,}045 \\
Prefix queries issued             & 2{,}051{,}801 \\
Tags resolved                     & 6{,}399{,}608 \\
Image digests resolved            & 7{,}416{,}671 \\
Layer nodes                 & 84.7\,M \\
\texttt{IS\_BASE\_OF} edges & 54{,}382{,}383 \\
Images in scan queue        & 6.7\,M \\
\texttt{last\_updated} coverage (crawl) & 95.7\% \\
\bottomrule
\end{tabular}
&
\begin{tabular}[t]{@{}lr@{}}
\multicolumn{2}{@{}l}{\textbf{Scan campaign (Stage~III)}}\\
\midrule
Repositories scanned          & \textbf{52{,}895} \\
Distinct images by content    & 51{,}751 \\
Scanners per image        & 6 \\
Worker machines           & 13 (Debian/Ubuntu) \\
Median scan time / image  & 117\,s \\
Distinct CVEs detected    & 52{,}624 \\
Per-image reports (DB)    & 192\,GB \\
MongoDB crawl             & 24.9\,GB \\
Neo4j layer graph         & 65.9\,GB \\
Total dataset             & \textbf{283\,GB} \\
\bottomrule
\end{tabular}
\end{tabular}

\vspace{6pt}

\begin{tabular}{@{}lrrcr@{}}
\multicolumn{5}{@{}l}{\textbf{Merged findings by category} \quad (52{,}895 repositories, total \textbf{170{,}373{,}044})}\\
\midrule
\textbf{Category} & \textbf{Findings} & \textbf{Share} & \textbf{Scanners} & \textbf{Img.\ $\geq$1} \\
\midrule
Package vulnerabilities      & 141{,}683{,}960 & 83.2\% & 3 & 96.3\% \\
Software-inventory components & 23{,}151{,}449  & 13.6\% & 1 & 96.3\% \\
Embedded secrets             & 4{,}955{,}892   & 2.9\%  & 2 & 76.9\% \\
Misconfigurations            & 581{,}743       & 0.3\%  & 2 & 98.0\% \\
\bottomrule
\end{tabular}
\end{table}

\section{Results}
\label{sec:results}

\label{sec:res-scale}
\sysname{} scanned the \textbf{52{,}895} highest-exposure repositories, a census of the \emph{most consequential} images on Docker Hub, not an upper bound over the whole registry. Exposure does not predict vulnerability (Section~\ref{sec:res-exposure}).

\subsection{Vulnerability Prevalence}
\label{sec:res-prevalence}
\begin{figure}[!htpb]
\centering
\includegraphics[width=\columnwidth]{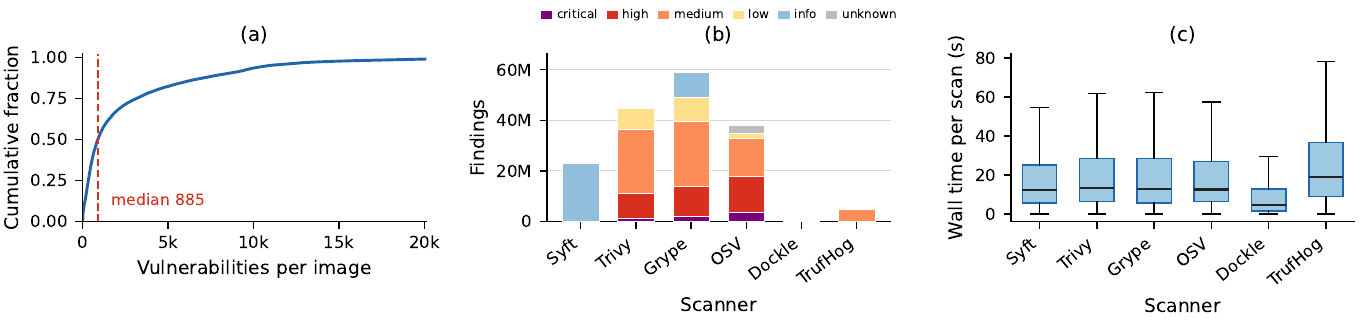}
\caption{Per-image: (a)~cumulative vulnerability distribution, (b)~by scanner/severity, (c)~scan time.}
\label{fig:results}
\end{figure}

Known vulnerabilities are pervasive. Of the scanned repositories, \textbf{96.3\%} carry at least one known package vulnerability (only 3.7\% are clean), \textbf{93.4\%} ship a critical-severity one and 95.5\% a high-severity one; the per-image count is heavily right-skewed (Figure~\ref{fig:results}(a)): the median image carries \textbf{885} merged findings, the mean is 2{,}679, the 99th percentile 20{,}109, and the maximum 113{,}166. These counts are merged, \emph{not} deduplicated across scanners, which inflates the mean (Section~\ref{sec:res-divergence}). Across the \textbf{141.7~million} package-vulnerability findings (Table~\ref{tab:severity}), 5.2\% are critical and 25.2\% high, a combined \textbf{43.1~million}; because a base image's layers are inherited by every image built on it (Section~\ref{sec:stage2}), each critical finding in a widely reused base propagates across its entire downstream subtree.

\begin{table}[!htpb]
\centering
\caption{Package-vulnerability findings by severity.}
\label{tab:severity}
\renewcommand{\arraystretch}{0.95}
\setlength{\tabcolsep}{4pt}
\footnotesize
% transposed: severities as columns to cut the vertical footprint in half
\begin{tabular*}{\columnwidth}{@{\extracolsep{\fill}}lrrrrrrr}
\toprule
 & \textbf{Critical} & \textbf{High} & \textbf{Medium} & \textbf{Low} & \textbf{Info.} & \textbf{Unrated} & \textbf{Total} \\
\midrule
Findings       & 7{,}377{,}355 & 35{,}740{,}343 & 65{,}727{,}713 & 19{,}418{,}002 & 9{,}927{,}589 & 3{,}492{,}958 & \textbf{141{,}683{,}960} \\
\% of findings & 5.2 & 25.2 & 46.4 & 13.7 & 7.0 & 2.5 & 100 \\
Images $\geq$1 & 49{,}392 & 50{,}534 & 50{,}872 & 49{,}951 & 33{,}708 & 37{,}393 & \textbf{50{,}957} \\
\% of images   & 93.4 & 95.5 & 96.2 & 94.4 & 63.7 & 70.7 & \textbf{96.3} \\
\bottomrule
\end{tabular*}
\end{table}

\subsection{Per-Scanner Output, Reliability and Cost}
\label{sec:res-perscanner}
The six scanners measure different things, and their volumes are not comparable across categories (Table~\ref{tab:perscanner}, Figure~\ref{fig:results}(b)): among the three vulnerability scanners Grype emits the most findings (59.1\,M), followed by Trivy (45.0\,M) and OSV-Scanner (38.0\,M). Syft inventories 23.2\,M software components, TruffleHog reports 4.9\,M secret findings and Dockle 0.24\,M misconfigurations. Reliability varies sharply. The share of images scanned without error was 94.6\% to 98.8\% for five scanners but only \textbf{79.9\% for Trivy} (10{,}614 invocations errored). The failures are specific to Trivy, not to the infrastructure. All six scanners run against the same locally exported image archive (Section~\ref{sec:stage3}), yet Syft and Grype error on only about 1{,}200 to 1{,}900 images each. Four in five of Trivy's error messages report a timeout acquiring the tool's local cache lock (``cache may be in use by another process''), a concurrency limitation of the scanner under parallel operation rather than a property of the images scanned. A measurement that relied on Trivy alone would therefore silently miss roughly one image in five.

\begin{table}[!htpb]
\centering
\caption{Findings by scanner and severity.}
\label{tab:perscanner}
\renewcommand{\arraystretch}{0.95}
\footnotesize
\setlength{\tabcolsep}{4pt}
\begin{tabular*}{\textwidth}{@{\extracolsep{\fill}}lrrrrrr}
\toprule
\textbf{Scanner} & \textbf{Findings} & \textbf{Critical} & \textbf{High} & \textbf{Medium} & \textbf{Low} & \textbf{Run OK} \\
\midrule
Syft       & 23{,}151{,}449 & --            & --             & --             & --            & 97.7\% \\
Trivy      & 44{,}957{,}561 & 1{,}478{,}679 & 9{,}813{,}224 & 25{,}275{,}077 & 8{,}069{,}525 & 79.9\% \\
Grype      & 59{,}061{,}336 & 2{,}163{,}552 & 11{,}733{,}733 & 25{,}802{,}146 & 9{,}354{,}135 & 96.4\% \\
OSV        & 38{,}046{,}373 & 3{,}741{,}595 & 14{,}304{,}383 & 14{,}738{,}700 & 2{,}169{,}974 & 94.6\% \\
Dockle     & 235{,}029      & 45{,}411      & --             & 42{,}587       & 147{,}031     & 98.0\% \\
TruffleHog & 4{,}921{,}296  & 2{,}209       & 0              & 4{,}919{,}087 & 0             & 98.8\% \\
\bottomrule
\end{tabular*}
\end{table}

\begin{figure}[!htpb]
\centering
\includegraphics[width=\columnwidth]{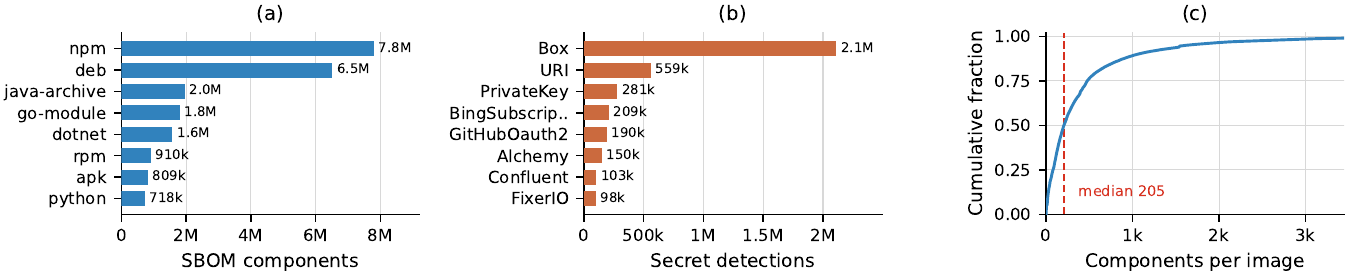}
\caption{(a)~Components by ecosystem, (b)~secrets by detector, (c)~components per image.}
\label{fig:inventory}
\end{figure}

The measurement is also costly. Grype is the slowest at 35.1\,s per image, and running all six on one image took a median of 117\,s (Figure~\ref{fig:results}(c)); the mean of 197\,s is inflated by a tail reaching 9{,}433\,s. Syft found a median of 205 components per image (mean 438, maximum 51{,}021; Figure~\ref{fig:inventory}(c)), dominated by \texttt{npm} (33.7\%) and \texttt{deb} (28.2\%), then Java, Go and .NET (Figure~\ref{fig:inventory}(a)).

\subsection{Inter-Scanner Divergence}
\label{sec:res-divergence}
\begin{figure}[!htpb]
\centering
\includegraphics[width=\columnwidth]{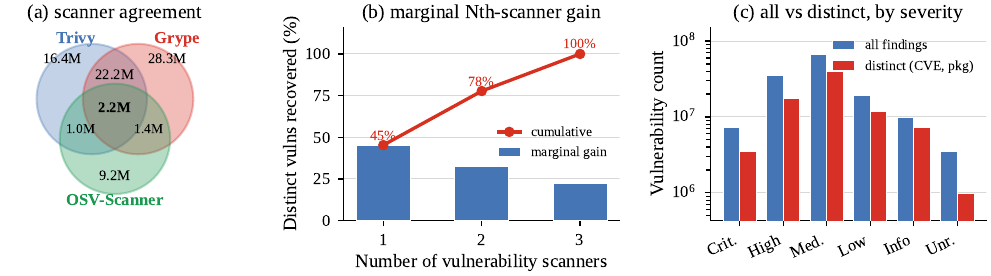}
\caption{Inter-scanner divergence: (a)~groups by subset, (b)~$N$th-scanner gain, (c)~all vs.\ distinct by severity.}
\label{fig:divergence}
\end{figure}

Running three vulnerability scanners on every image lets us quantify how much the measurement depends on the tool. We group the package-vulnerability findings within each image by \mbox{(vulnerability, package)}. Of the \textbf{80.7~million} distinct groups, \textbf{66.8\%} are reported by only one scanner and just \textbf{2.7\%} by all three (Figure~\ref{fig:divergence}(a)); even the closest pair, Grype and Trivy, co-detect only 22.2\,M. The divergence also shows in the aggregate counts. Grype reports 59.1\,M findings against OSV's 38.0\,M, a \textbf{1.55$\times$} spread (Table~\ref{tab:perscanner}). The best single scanner recovers only \textbf{66.9\%} of distinct groups (Figure~\ref{fig:divergence}(a)), so a one-tool measurement misses about one distinct vulnerability in three. Figure~\ref{fig:divergence}(b) shows the marginal gain of each added scanner, and Figure~\ref{fig:divergence}(c) contrasts merged findings with distinct groups per severity. This is consistent with the small-sample scanner-agreement literature~\cite{kaur2021scientific,mills2023longitudinal}, here at the scale of our full corpus: a single Docker Hub vulnerability count reflects, in large part, the scanner used to obtain it.

\subsection{Secrets and Misconfiguration}
\label{sec:res-secrets}
Beyond vulnerabilities, \sysname{} measures two further dimensions. Dockle flagged at least one misconfiguration from the CIS Docker Benchmark in \textbf{98.0\%} of images (51{,}853 of them); the most frequent (Table~\ref{tab:misconfig}) are a missing content-trust configuration, a missing \texttt{HEALTHCHECK} instruction, and the container running as \texttt{root}. Of Dockle's findings, 45{,}411 are at its \texttt{FATAL} checkpoint level, which we treat as critical severity. These flag concrete security faults: the most consequential is a credential stored in an environment variable or file (check CIS-DI-0010), which Dockle raised for \textbf{10{,}303} images, and others include an empty user password and the presence of \texttt{sudo} in the image. These counts rest on structural properties of the image rather than on verdicts from a vulnerability database, so they are less tied to any single scanner's vulnerability feed, though we do not measure their scanner-to-scanner stability directly. The near-universal prevalence is in line with studies of Dockerfile authoring practice, which find Dockerfile smells in a large share of images~\cite{durieux2024smellsize}.

TruffleHog flagged at least one embedded secret in \textbf{76.9\%} of images (40{,}667 images, 4.9\,M secret findings). This raw prevalence must be read with care. It counts \emph{detector hits}, that is, strings matching a detection pattern, not exploitable credentials. As Dr.\ Docker and Dahlmanns et al.~\cite{shi2025drdocker,dahlmanns2023secrets} document, the large majority of TruffleHog hits are example keys, test fixtures, or already-public material; Dr.\ Docker found 99.3\% of its raw detections invalid after filtering. We confirm this directly on our corpus. We drew a uniform random sample of \textbf{1{,}100} detections (95\% confidence, $\pm$3\% margin) and classified \emph{every one} by hand, rather than only the residual left after rule-based filtering. Just \textbf{three} were plausible live credentials: an application service-account key, an SSH host key, and an access token captured in an application log. The other \textbf{1{,}097} (\textbf{99.7\%}, Wilson 95\% CI 99.2 to 99.9\%) are non-credentials. About 60\% are package checksums and OS metadata; the rest are test fixtures, dependency caches, documentation examples, placeholders, and binary or locale artifacts. Every \texttt{PrivateKey} detection but the service-account key fell in a crypto-library self-test, an SSH/TLS test fixture, or a shared object. The 76.9\% detector-hit prevalence is therefore an upper bound dominated by false positives: only about 0.3\% of detections are plausible credentials, so the prevalence of \emph{validated} secrets is far lower. Active validation of these residual candidates is left to future work (Section~\ref{sec:conclusion}). The detector mix (Figure~\ref{fig:inventory}(b)) corroborates this. Generic URI and private-key patterns dominate, exactly the categories most prone to false positives from example values.

Beyond prevalence, the \emph{distribution} of how many secrets an image carries is itself heavy-tailed (Figure~\ref{fig:secretcdf}). The median image carries 13 detector hits, but the 99th percentile reaches 1{,}230 and the image with the most hits reports 19{,}856. Official images, although slightly more likely to carry at least one secret (Section~\ref{sec:res-official}), carry far fewer of them; their per-image count tops out at 1{,}251 against the community tail above 19{,}000. This is consistent with Dahlmanns et al.~\cite{dahlmanns2023secrets}, who likewise found that a minority of images account for most embedded secrets; here the distribution makes that concentration explicit.

% Dockle misconfiguration table and the secret CDF figure share one float,
% side by side: both belong to the Secrets and Misconfiguration subsection and
% the pairing saves vertical space. Both minipages are top-aligned ([t]) so the
% two captions start at the same height and the pair reads as aligned; the
% tabular keeps the table counter and the figure uses \captionof for the
% figure counter.
\begin{table}[!htpb]
\centering
\begin{minipage}[t]{0.52\columnwidth}
\captionsetup{justification=raggedright,singlelinecheck=false}
\caption{Top Dockle misconfiguration checks.}
\label{tab:misconfig}
\footnotesize
\setlength{\tabcolsep}{2.6pt}
\begin{tabular*}{\linewidth}{@{\extracolsep{\fill}}lrr}
\toprule
\textbf{Misconfiguration check} & \textbf{Images} & \textbf{\%} \\
\midrule
Content-trust off (CIS-DI-0005)        & 51{,}847 & 98.0 \\
No \texttt{HEALTHCHECK} (CIS-DI-0006)  & 49{,}792 & 94.1 \\
No dedicated user (CIS-DI-0001)        & 41{,}863 & 79.1 \\
\texttt{setuid}/\texttt{setgid} files (CIS-DI-0008) & 31{,}785 & 60.1 \\
Credential in env/file (CIS-DI-0010)   & 10{,}303 & 19.5 \\
Empty user password (DKL-LI-0001)      & 4{,}867 & 9.2 \\
\texttt{sudo} present (DKL-DI-0001)    & 4{,}168 & 7.9 \\
\bottomrule
\end{tabular*}
\end{minipage}\hfill
\begin{minipage}[t]{0.45\columnwidth}
\centering
\captionsetup{justification=raggedright,singlelinecheck=false}
% caption on top, matching the table caption, so the two minipage tops align
\captionof{figure}{Secret detections per image (CDF): official vs.\ community.}
\label{fig:secretcdf}
\vspace{2pt}
\includegraphics[width=\linewidth]{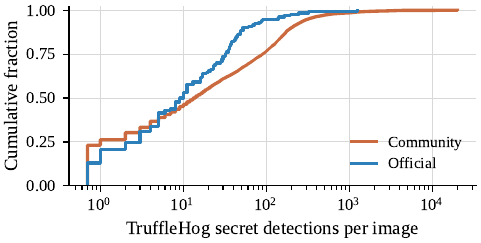}
\end{minipage}
\end{table}

\subsection{Official versus Community Images}
\label{sec:res-official}
\begin{figure}[!htpb]
\centering
\includegraphics[width=0.80\columnwidth]{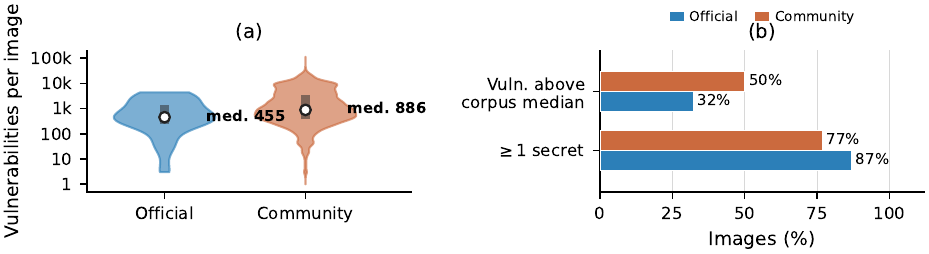}
\caption{Official vs.\ community: (a)~vulnerabilities per image, (b)~secret prevalence and share above median.}
\label{fig:offcomm}
\end{figure}

Splitting the corpus by namespace (official images occupy \texttt{library}) reproduces Liu et al.'s~\cite{liu2020understanding} official-versus-community comparison (Figure~\ref{fig:offcomm}(a)). The 130 official images carry a median of \textbf{455} merged vulnerability findings, the 52{,}765 community images \textbf{886} (means: 1{,}046 official, 2{,}683 community), and 50.0\% of community images sit above the corpus median against 32.3\% of official ones: community images remain the more vulnerable population, as Liu reported, though both now sit far higher in absolute counts. Secrets show the reverse ordering (Figure~\ref{fig:offcomm}(b)): TruffleHog flags a secret in \textbf{86.9\%} of official against \textbf{76.9\%} of community images, but these are raw hits and the gap is modest. The same ordering held in Shu et al.~\cite{shu2017dockerhub} (community median 158 in 2017), so it persists across nine years even as counts rose.

\subsection{Popularity, Exposure and Security Posture}
\label{sec:res-exposure}
\begin{figure}[!htpb]
\centering
\includegraphics[width=\columnwidth]{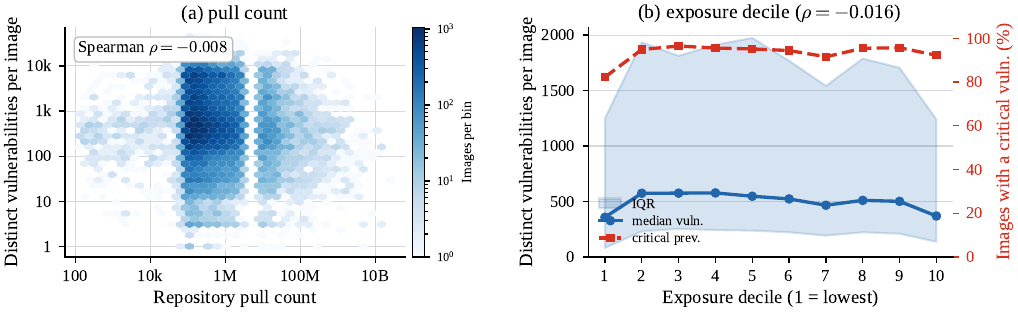}
\caption{(a)~Vulnerabilities vs.\ pull count; (b)~by exposure decile.}
\label{fig:exposure}
\end{figure}

A natural question for a popularity-ranked corpus is whether the images the ecosystem most depends on are better maintained. They are not. Raw pull count is essentially uncorrelated with distinct vulnerability count (Spearman rank correlation \textbf{$\rho=-0.01$}, Figure~\ref{fig:exposure}(a)); our exposure metric (Section~\ref{sec:exposure}), which adds the pulls flowing through an image's downstream layer subtree, gives the same verdict. Across exposure deciles (Figure~\ref{fig:exposure}(b)) the median vulnerability count shows no clear trend (\textbf{369} in the highest-exposure decile), with \textbf{$\rho=-0.02$} for exposure against vulnerability count and \textbf{$\rho=-0.04$} against \emph{critical}-vulnerability count, both negligible. Critical prevalence stays near 93\% across the higher-exposure deciles, so even the most heavily reused base images almost all ship a critical finding, echoing a Maven Central study where dependency centrality is not a safeguard against vulnerability~\cite{haq2025ripple}.

\subsection{Exposure-Weighted Vulnerability Risk}
\label{sec:res-reach}
Counting the images a CVE affects treats a flaw in \texttt{alpine} and a flaw in an abandoned image with no dependents as equal. Because every image in the corpus carries an exposure score, we can instead weight each CVE by the ecosystem reach of the images it affects. We sum \texttt{exposure} over the images that carry each CVE, which ranks vulnerabilities by the share of the ecosystem they can affect rather than by raw image count. In Table~\ref{tab:reach}, ``\% expo.'' is the share of total corpus exposure carried by the affected images. The ranking is dominated by recent CVEs in \texttt{zlib} and \texttt{glibc}, the C libraries embedded in nearly every base image, together with the Go standard library (\texttt{stdlib}) that ships inside the many Go-based images in the corpus: the single highest-reach CVE touches images accounting for \textbf{47.3\%} of the corpus's total exposure while being present in only 44.9\% of images by count. This gap between image count and exposure share is the central finding: a CVE in a base library is not just one of many; it endangers a share of the ecosystem far larger than its image count suggests, and the exposure metric makes that quantitatively concrete.

\begin{table}[!htpb]
\centering
\caption{Highest-reach CVEs by exposure.}
\label{tab:reach}
\footnotesize
\setlength{\tabcolsep}{4pt}
\renewcommand{\arraystretch}{0.95}
\begin{tabular}{llrrr}
\toprule
\textbf{CVE} & \textbf{Package} & \textbf{Sev.} & \textbf{Images} & \textbf{\% expo.} \\
\midrule
CVE-2026-27171 & \texttt{zlib} & High & 23{,}726 & 47.3\% \\
CVE-2026-4046 & \texttt{glibc} & High & 18{,}774 & 46.5\% \\
CVE-2026-5928 & \texttt{glibc} & High & 18{,}599 & 46.2\% \\
CVE-2026-5435 & \texttt{glibc} & Crit. & 18{,}549 & 46.1\% \\
CVE-2026-5450 & \texttt{glibc} & Crit. & 18{,}086 & 45.6\% \\
CVE-2026-33814 & \texttt{stdlib} & High & 12{,}873 & 44.6\% \\
CVE-2026-33811 & \texttt{stdlib} & High & 12{,}778 & 43.2\% \\
CVE-2026-39820 & \texttt{stdlib} & High & 12{,}778 & 43.2\% \\
\bottomrule
\end{tabular}
\end{table}

The layer graph lets us go one step further and measure how far a vulnerability \emph{propagates}. For each CVE we count the \emph{distinct} downstream images that inherit a vulnerable layer. Starting from the images that directly carry the CVE, we take the union of their descendant subtrees in the \texttt{IS\_BASE\_OF} forest. Each distinct downstream image is counted once (Table~\ref{tab:propagation}). Taking the union, rather than summing each affected image's downstream count, avoids double-counting the descendants that several affected base images share. The reach is nonetheless large. Each top CVE is inherited by \textbf{1.0} to \textbf{1.14~million} distinct downstream images, a propagation factor (distinct downstream over directly affected images) of \textbf{39} to \textbf{49}, with a single \texttt{zlib} flaw reaching \textbf{1{,}129{,}391}. These counts quantify, on the reconstructed \texttt{IS\_BASE\_OF} forest, how far a single flaw spreads through the supply chain, and they render per-CVE the parent-to-child vulnerability inheritance that Shu et al.~\cite{shu2017dockerhub} first documented at the level of whole images. The top-reach CVEs sit in libraries that ship with nearly every base image (\texttt{zlib}, \texttt{glibc}, \texttt{libcrypto3}), so part of their reach reflects sheer ubiquity rather than the layer hierarchy; the columns of Table~\ref{tab:propagation} separate these effects, with the direct count measuring how widespread the vulnerable library is among the scanned images and the propagation factor measuring the amplification contributed by layer inheritance alone. Direct counts are restricted to images resolved in the Stage~II layer graph, so they sit slightly below the affected-image counts of Table~\ref{tab:reach}. One caveat remains. The \texttt{IS\_BASE\_OF} graph spans only the Stage~II-resolved fraction of the crawl, so these distinct-downstream counts are structural \emph{lower} bounds; the true reach on Docker Hub is at least this large.

\begin{table}[!htpb]
\centering
\caption{Per-CVE downstream propagation.}
\label{tab:propagation}
\small
\setlength{\tabcolsep}{4pt}
\begin{tabular}{llrrr}
\toprule
\textbf{CVE} & \textbf{Package} & \textbf{Direct} & \textbf{Distinct downstr.} & \textbf{Factor} \\
\midrule
CVE-2026-28388 & \texttt{libcrypto3} & 28{,}343 & 1{,}144{,}010 & 40.4 \\
CVE-2026-28390 & \texttt{libcrypto3} & 27{,}768 & 1{,}134{,}329 & 40.9 \\
CVE-2026-28389 & \texttt{libcrypto3} & 27{,}758 & 1{,}134{,}319 & 40.9 \\
CVE-2026-27171 & \texttt{zlib} & 23{,}013 & 1{,}129{,}391 & 49.1 \\
CVE-2026-28387 & \texttt{libcrypto3} & 23{,}819 & 1{,}096{,}284 & 46.0 \\
CVE-2026-22796 & \texttt{libcrypto3} & 26{,}279 & 1{,}033{,}461 & 39.3 \\
CVE-2025-68160 & \texttt{libcrypto3} & 26{,}279 & 1{,}033{,}461 & 39.3 \\
CVE-2025-69421 & \texttt{libcrypto3} & 26{,}278 & 1{,}033{,}459 & 39.3 \\
\bottomrule
\end{tabular}
\end{table}

\subsection{Comparison with Prior Measurements}
\label{sec:res-comparison}
\begin{figure}[!htpb]
\centering
\includegraphics[width=\columnwidth]{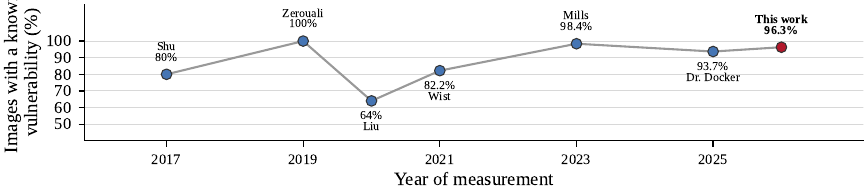}
\caption{Known-vulnerability prevalence across studies, 2017 to 2026.}
\label{fig:timeline}
\end{figure}

Absolute vulnerability counts are not comparable across studies, since the measurements differ in scanner, sample and elapsed time, and our own data shows a 1.55$\times$ inter-scanner spread in finding volume alone (Section~\ref{sec:res-divergence}); we therefore compare prevalence and orders of magnitude rather than raw counts. Known-vulnerability prevalence shows continuity, not reversal. Across 2017 to 2025 it has stayed high (Shu: $>$80\% of images; Liu: $>$64\% of community images; Wist: 81.8\% of certified images; Mills: 98.4\% of images, 374 of 380; Dr.\ Docker: 93.7\% of images; each with a different metric and scanner). Our \textbf{96.3\%} is at the upper end of this range, on par with the most recent multi-scanner study (Mills, 98.4\%), so prevalence has not declined over the eight years these studies span (Figure~\ref{fig:timeline}).

\subsection{Reproducing Prior Docker Hub Analyses}
\label{sec:res-repro}
\begin{figure}[!htpb]
\centering
\includegraphics[width=0.86\columnwidth]{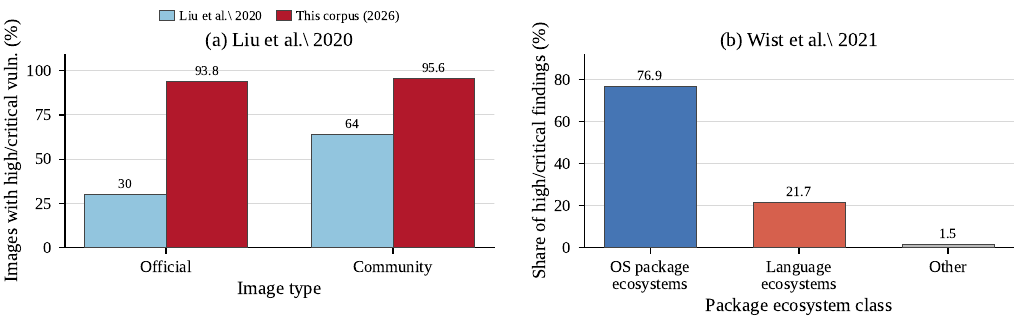}
\caption{Reproductions: (a)~official vs.\ community prevalence~\cite{liu2020understanding}; (b)~severe findings by ecosystem~\cite{wist2021vulnerability}.}
\label{fig:repro}
\end{figure}

\begin{figure}[!htpb]
\centering
\includegraphics[width=\columnwidth]{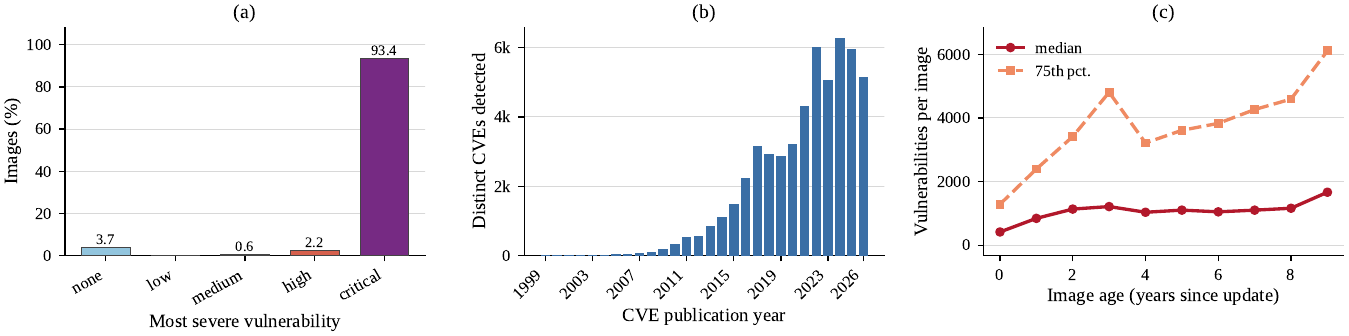}
\caption{Shu et al.~\cite{shu2017dockerhub} reproduced: (a)~worst severity, (b)~CVEs by year, (c)~vulns vs.\ image age.}
\label{fig:shu}
\end{figure}

To ground the comparison in method, we reproduced one concrete analysis from each of six prior Docker Hub measurements on our corpus (Table~\ref{tab:repro}; Figures~\ref{fig:repro} and~\ref{fig:shu}). As above, these comparisons show trends, not exact replications.

\begin{table}[!htpb]
\centering
\caption{Prior measurements reproduced on our corpus.}
\label{tab:repro}
\renewcommand{\arraystretch}{1.0}
\scriptsize
\setlength{\tabcolsep}{3pt}
\begin{tabular}{@{}>{\raggedright\arraybackslash}p{0.205\columnwidth} >{\raggedright\arraybackslash}p{0.255\columnwidth} >{\raggedright\arraybackslash}p{0.235\columnwidth} >{\raggedright\arraybackslash}p{0.235\columnwidth}@{}}
\toprule
\textbf{Study} & \textbf{Reproduced analysis} & \textbf{Reported by the study} & \textbf{This corpus} \\
\midrule
Shu et al.~\cite{shu2017dockerhub}        & Worst-severity bucket; community vuln.\ median & \emph{high} modal; community median 158        & \emph{critical} modal (93.4\%); median 885 \\
Zerouali et al.~\cite{zerouali2019outdated} & Vulnerabilities-per-image distribution        & median 601, mean 1{,}336, max 7{,}338          & median 885, mean 2{,}679, max 113{,}166 \\
Liu et al.~\cite{liu2020understanding}     & High/critical prevalence, official vs.\ comm.\ & ${\sim}$30\% official, ${>}$64\% community      & 93.8\% official, 95.6\% community \\
Wist et al.~\cite{wist2021vulnerability}   & Severe findings by ecosystem (OS vs.\ language) & severe surface in language ecosystems          & 76.9\% OS, 21.7\% language \\
Mills et al.~\cite{mills2023longitudinal}  & Oldest CVE still present; staleness effect     & CVEs back to 1999                              & CVEs back to 1999; median 414 (recent) vs.\ 1{,}085 (old images) \\
Dahlmanns et al.~\cite{dahlmanns2023secrets} & Secret prevalence; private-key category      & 8.5\% validated; private keys dominant         & 76.9\% hits, 99.7\% FP (hand-labeled) \\
\bottomrule
\end{tabular}
\end{table}

\paragraph{Shu, Zerouali, Liu.} Shu et al.~\cite{shu2017dockerhub} classified images by worst severity, ranked vulnerable packages, and tracked CVEs by year (Figure~\ref{fig:shu}): the most common (modal) worst severity has shifted from \emph{high} to \emph{critical} (Figure~\ref{fig:shu}(a)); the \textbf{52{,}624} distinct CVEs reach back to 1999 but \textbf{68.4\%} are 2020-or-later (Figure~\ref{fig:shu}(b)); and the top vulnerable packages remain operating-system libraries (\texttt{zlib} 44{,}574 images, \texttt{openssl} 39{,}209; Table~\ref{tab:toppkg}). Zerouali et al.~\cite{zerouali2019outdated} found essentially every Debian-based container affected; we reproduce this at \textbf{96.3\%} prevalence. Liu et al.~\cite{liu2020understanding} split official versus community (Figure~\ref{fig:repro}(a)): the community-above-official ordering holds, but both populations now sit far higher.

\paragraph{Wist, Mills, Dahlmanns.} Wist et al.~\cite{wist2021vulnerability} found severe vulnerabilities concentrated in language ecosystems; our corpus, ranked by exposure and dominated by OS images, inverts this (Figure~\ref{fig:repro}(b)): of \textbf{43.1~million} severe findings most are operating-system packages, with Go (5.3~M), \texttt{npm} (1.3~M) and Python (1.0~M) leading the language share. Mills et al.~\cite{mills2023longitudinal} noted CVEs back to 1999 and a staleness effect; both patterns hold in our corpus: images updated within a year carry a median \textbf{414} vulnerabilities against \textbf{1{,}085} for older ones (Figure~\ref{fig:shu}(c)), consistent with the rising vulnerability lifespans reported across open-source ecosystems~\cite{akhavani2025opensource}. Dahlmanns et al.~\cite{dahlmanns2023secrets} found private keys dominant (8.5\% validated); TruffleHog flags a private-key pattern in \textbf{40.2\%} of images, but our hand-labeling puts its false-positive rate at \textbf{99.7\%} (Section~\ref{sec:res-secrets}).

% Compacted in place: footnotesize body and tight rows. Six numeric columns
% make this table too wide to pair side by side with a figure.
\begin{table}[!htpb]
\centering
\caption{Ten most prevalent vulnerable packages.}
\label{tab:toppkg}
\footnotesize
\setlength{\tabcolsep}{4pt}
\renewcommand{\arraystretch}{0.9}
\begin{tabular}{lrrrrr}
\toprule
\textbf{Package} & \textbf{Vuln.\ img.} & \textbf{\% corpus} & \textbf{Findings} & \textbf{CVEs} & \textbf{Img.\ w/ crit.} \\
\midrule
\texttt{zlib} & 44{,}574 & 84.3\% & 177{,}503 & 15 & 28{,}695 \\
\texttt{openssl} & 39{,}209 & 74.1\% & 2{,}597{,}921 & 290 & 27{,}479 \\
\texttt{tar} & 32{,}031 & 60.6\% & 499{,}294 & 27 & 0 \\
\texttt{util-linux} & 28{,}555 & 54.0\% & 785{,}349 & 21 & 14 \\
\texttt{curl} & 28{,}516 & 53.9\% & 2{,}836{,}304 & 187 & 18{,}309 \\
\texttt{coreutils} & 27{,}896 & 52.7\% & 142{,}134 & 33 & 0 \\
\texttt{ncurses-base} & 27{,}671 & 52.3\% & 372{,}917 & 30 & 1{,}440 \\
\texttt{systemd} & 27{,}343 & 51.7\% & 1{,}121{,}059 & 57 & 3{,}476 \\
\texttt{expat} & 25{,}886 & 48.9\% & 649{,}552 & 47 & 16{,}282 \\
\texttt{bash} & 25{,}857 & 48.9\% & 85{,}097 & 15 & 19 \\
\bottomrule
\end{tabular}
\end{table}

\section{Limitations and Conclusion}
\label{sec:conclusion}

\noindent\textbf{Limitations.} The corpus is \emph{not} a random sample but the top of the exposure ranking, so prevalences describe the most consequential images, not the average one, and are not an upper bound (exposure does not predict vulnerability). The 76.9\% secret figure counts raw detector hits (99.7\% non-credentials by hand-labeling), and reported vulnerabilities are likewise static scanner findings, not confirmed exploitable. Per-image counts are merged, not deduplicated (we report medians), and the unit is one \texttt{latest} image per repository (deduplicating to \textbf{51{,}751} digests leaves results unchanged). We scanned \texttt{linux/amd64} only, and Stage~II resolved only the most popular repositories (44\%) and recent tags, so the propagation counts are \emph{lower bounds}. Finally, the scan-reports database is frozen, but the crawl (MongoDB) and layer graph (Neo4j) are snapshots of a pipeline that kept running past the freeze, so the released copies sit a few percent above the counts reported here; every scan-derived result reproduces exactly from the frozen database, and the crawl- and graph-derived counts reproduce to within that margin.

\noindent\textbf{Practical implications.} Three practices follow from the measurements. First, because the best single scanner recovers only 66.9\% of distinct vulnerabilities (Section~\ref{sec:res-divergence}), audits and continuous-integration (CI) pipelines should run at least two vulnerability scanners backed by different databases rather than trust a single count. Second, exposure gives registry operators and platform teams a remediation order. It identifies the base images whose flaws reach the most pulls, so addressing those first targets the largest share of downstream inheritance, whereas ranking flaws by raw affected-image count alone would spend the same effort on images the ecosystem barely pulls (Section~\ref{sec:res-reach}). Third, raw secret detections should not gate a deployment. With 99.7\% of hits being non-credentials (Section~\ref{sec:res-secrets}), a validation step must precede any action on them, while the structural misconfigurations Dockle flags (running as \texttt{root}, missing \texttt{HEALTHCHECK}) are cheap to check and enforce in CI.

\noindent\textbf{Conclusion.} We measured the security posture of the highest-exposure Docker Hub images,\footnote{Code and dataset: \url{https://github.com/ChimangoScan/chimangoscan}} the bases on which 84.7\% of all recorded pulls depend, with six independent scanners rather than one. Vulnerabilities, critical ones, and misconfigurations are near-universal, yet the single number a lone tool reports is largely a property of that tool, so ecosystem-scale measurements should report a scanner battery, not a single count. Exposure marks where a flaw does the most damage yet does not predict how vulnerable an image is; resolving Stage~II fully and validating the residual secrets are natural extensions. We release the pipeline and dataset for audit and reuse.\footnote{Generative AI tools assisted with writing and language revision of this manuscript; the authors reviewed all content and take full responsibility for it.}

%
% ---- Bibliography ----
%
\bibliographystyle{plain}
\bibliography{paper}

\end{document}